\documentstyle[preprint,pra,aps,fleqn]{revtex}

\begin{document}
\draft
\title{Nonclassical properties and algebraic characteristics of negative binomial
states in quantized radiation fields}
\author{Xiao-Guang Wang\thanks{%
email: xyw@aphy.iphy.ac.cn\newline
Ref.number: D9123\newline
Section: Optical Physics and Quantum Optics\newline
Running title: Negative binomial states in quantized radiation fields\newline
Accepted in EPJD on 22/11/99
} Shao-Hua Pan and Guo-Zhen Yang}
\address{CCAST(World Laboratory),P.O.Box 8730, Beijing 100080 \\
and Laboratory of Optical Physics, Institute of Physics,Chinese Academy of\\
Sciences,Beijing 100080, P.R.China}
\date{\today}
\maketitle

\begin{abstract}
We study the nonclassical properties and algebraic characteristics of the
negative binomial states introduced by Barnett recently. The ladder operator
formalism and displacement operator formalism of the negative binomial
states are found and the algebra involved turns out to be the SU(1,1) Lie
algebra via the generalized Holstein-Primarkoff realization. These states
are essentially Peremolov's SU(1,1) coherent states. We reveal their
connection with the geometric states and find that they are excited
geometric states. As intermediate states, they interpolate between the
number states and geometric states. We also point out that they can be
recognized as the nonlinear coherent states. Their nonclassical properties,
such as sub-Poissonian distribution and squeezing effect are discussed. The
quasiprobability distributions in phase space, namely the Q and Wigner
functions, are studied in detail. We also propose two methods of generation
of the negative binomial states.
\end{abstract}


\section{Introduction}

Since Stoler et al. introduced the binomial states (BSs)[1], the so-called
intermediate states have attracted considerable attention of physicists in
the field of quantum optics. A feature of these states is that they
interpolate between two fundamental quantum states, such as the number,
coherent and squeezed states, and reduce to them in two different limits.
For instance, the BSs interpolate between the coherent states (the most
classical) and the number states (the most nonclassical)[1-6], while the
negative binomial states (NBSs) interpolate between the coherent states and
geometric states[7-11]. Another feature of some intermediate states is that
their photon number distributions are some famous discrete probability
distributions in probability theory: the BS corresponds to the binomial
distribution, the NBS to the negative binomial distribution, the
hypergeometric state[12] to the hypergeometric distribution, and the
negative hypergeometric state[13] to the negative hypergeometric
distribution.

Recently Barnett introduced a new definition of NBS[14], 
\begin{eqnarray}
|\eta ,M\rangle &=&\sum_{n=M}^\infty C_n(\eta ,M)|n\rangle  \nonumber \\
&=&\sum_{n=M}^\infty \left[ {%
{{n} \choose {M}}%
}\eta ^{M+1}(1-\eta )^{n-M}\right] ^{1/2}|n\rangle ,
\end{eqnarray}
where $|n\rangle $ is the usual number state, $0<\eta \le 1$ and $M$ is a
non-negative integer. They find that the NBS $|\eta ,M\rangle $ and the BS
have similar properties if the roles of the creation operator $a^{\dagger }$
and annihilation operator $a$ are interchanged. The photon number
probability $|C_n^{}(\eta ,M)|^2$ is associated with the probability that $n$
photons were present given that $M$ are found and that the probability for
successfully detecting any single photon is $\eta $. Mixed states with the
photon number probability include those applicable to photodetection and
optical amplification[15].

The BS is a intermediate number-coherent state and the original NBS is a
intermediate geometric-coherent state. One question naturally arises that if
there exist an intermediate state which interpolates between the number and
geometric state. In reality, the new NBS is just the intermediate
number-geometric state. This fact will be seen in the next section. Thus, we
have three intermediate states which interpolate between two of the three
fundamental states (the number, coherent and geometric states).

In the present paper we shall study the nonclassical properties and
algebraic characteristics of the new NBS. The ladder operator formalism ,
displacement operator formalism and related algebraic structure will be
formulated in section II. It is interesting that the algebraic structure is
the SU(1,1) Lie algebra via the generalized Holstein-Primarkoff realization.
In section III, we will show that the NBS can be viewed as excited geometric
states, intermediate number-geometric states and nonlinear coherent states.
The nonclassical properties, such as sub-Poissonian distribution and
squeezing effect will be investigated in detail in section IV. The Q and
Wigner functions are studied in Section V and the two methods of generation
of the NBS are proposed in section VI. A conclusion is given in section VII.

\section{Ladder operator formalism ,displacement operator formalism and
algebraic structure of the new NBS}

\subsection{Ladder operator formalism and algebraic structure}

It is known that the BSs are special SU(2) coherent states[5,6] and the
original NBSs are Perelomov's SU(1,1) coherent states[11] via the standard
Holstein-Primakoff realizations. So we expect that the algebra involved in
the new NBS is SU(1,1) Lie algebra.

It is easy to evaluate that 
\begin{equation}
a^{\dagger n}|\eta ,M\rangle =\left[ \frac{(M+n)!}{M!\eta ^n}\right]
^{1/2}|\eta ,M+n\rangle .
\end{equation}
In particular, for $n=1$, we get 
\begin{equation}
a^{\dagger }|\eta ,M\rangle =\left( \frac{M+1}\eta \right) ^{1/2}|\eta
,M+1\rangle .
\end{equation}
The creation operator raises the NBS $|\eta ,M\rangle $ to $|\eta
,M+1\rangle $. This property is similar to the action of the creation
operator on the Fock state $|M\rangle $, 
\begin{equation}
a^{\dagger }|M\rangle =\sqrt{M+1}|M+1\rangle .
\end{equation}
Actually, in the limit of $\eta \rightarrow 1$, the NBS $|\eta ,M\rangle
^{-} $ reduces to the number state $|M\rangle $ and Eq.(3) naturally reduces
to Eq.(4).

The key and interesting point is that there exists another operator $\sqrt{%
\hat{N}-M}$ which also raises the NBS $|\eta ,M\rangle $ to $|\eta
,M+1\rangle $. From Eq.(1), the following equation is directly derived as 
\begin{equation}
\sqrt{\hat{N}-M}|\eta ,M\rangle =\left( {\frac{1-\eta }\eta }\right) ^{1/2}%
\sqrt{M+1}|\eta ,M+1\rangle .
\end{equation}
Comparing Eq.(3) and Eq.(5), we get 
\begin{equation}
\sqrt{\hat{N}-M}|\eta ,M\rangle =\sqrt{1-\eta }a^{\dagger }|\eta ,M\rangle .
\end{equation}
Multiplying the both sides of the above equation by the operator $\sqrt{\hat{%
N}-M}$ from left, we obtain the ladder operator formalism of the NBS as 
\begin{equation}
(\hat{N}-\sqrt{1-\eta }\sqrt{\hat{N}-M}a^{\dagger })|\eta ,M\rangle =M|\eta
,M\rangle .
\end{equation}
In the limit of $\eta \rightarrow 1$, we find that the NBS$|\eta ,M\rangle $
reduces to the number state $|M\rangle $ and Eq.(7) to the equation $\hat{N}%
|M\rangle =M|M\rangle $ as expected.

It can be proved that the operators appearing in Eq.(7) can form SU(1,1) Lie
algebra: 
\begin{equation}
K_0=\hat{N}-\frac{M-1}2,K_{+}=\sqrt{\hat{N}-M}a^{\dagger },K_{-}=a\sqrt{\hat{%
N}-M}.
\end{equation}
Reminding of the standard Holstein-Primarkoff realization of SU(1,1) Lie
algebra, $J_0=\hat{N}+\frac M2,J_{+}=\sqrt{M+\hat{N}-1}a^{\dagger },J_{-}=a%
\sqrt{M+\hat{N}-1}$, we call the new realization as generalized
Holstein-Primakoff realization. To our knowledge, the new realization of
SU(1,1) Lie algebra seems not to be addressed in the literature.

In terms of the generators $K_0,K_{+}$ and $K_{-}$ of the SU(1,1) Lie
algebra , Eq.(7) is rewritten as 
\begin{equation}
(K_0-\sqrt{1-\eta }K_{+})|\eta ,M\rangle =\frac{M+1}2|\eta ,M\rangle .
\end{equation}

\subsection{Displacement operator formalism}

Now we try to find the displacement operator formalism of the NBS. To this
end, let us rewrite the NBS as 
\begin{eqnarray}
|\eta ,M\rangle &=&\eta ^{\frac{M+1}2}\sum_{n=0}^\infty \left[ {{%
{{M+n} \choose {M}}%
}}(1-\eta )^n\right] ^{1/2}|M+n\rangle  \nonumber \\
&=&\eta ^{\frac{M+1}2}\sum_{n=0}^\infty \frac{(\sqrt{1-\eta }a^{\dagger })^n%
}{\sqrt{n!}}|M\rangle .
\end{eqnarray}
Then by making use of the following identity 
\begin{equation}
\lbrack f(\hat{N})a^{\dagger }]^n=a^{\dagger n}f(\hat{N}+n)f(\hat{N}%
+n-1)...f(\hat{N}+1),
\end{equation}
Eq.(10) can be written in the exponential form 
\begin{equation}
|\eta ,M\rangle =\eta ^{\frac{M+1}2}e^{{\sqrt{1-\eta }K_{+}}}|M\rangle .
\end{equation}
Note that $K_{-}|M\rangle =0$, then the displacement operator formalism is
obtained as 
\begin{eqnarray}
|\eta ,M\rangle &=&e^{\sqrt{1-\eta }K_{+}}\eta ^{K_0}e^{-\sqrt{1-\eta }%
K_{-}}|M\rangle  \nonumber \\
&=&e^{\xi (K_{+}-K_{-})}|M\rangle ,
\end{eqnarray}
where $\xi $=arctanh $\sqrt{1-\eta }$. In the derivation of the above
equation, we have used the identity 
\begin{equation}
e^{\alpha K_{+}-\alpha ^{*}K_{-}}=e^{\gamma K_{+}}(1-|\gamma
|^2)^{K_0}e^{-\gamma ^{*}K_{-}},
\end{equation}
where $\gamma =\alpha \tanh |\alpha |/|\alpha |$. As seen from Eq.(13), the
NBS can be simply recognized as SU(1,1) displaced number states. Actually,
the NBS are essentially Peremolov's coherent states as shown below.

On the space 
\begin{equation}
S=\text{span}\left\{ {|n+M\rangle \equiv |n;k\rangle |n=0,1,2...}\right\} ,k=%
\frac{M+1}2
\end{equation}
we have 
\begin{eqnarray}
K_{+}|n;k\rangle &=&\sqrt{(n+1)(2k+n)}|n+1;k\rangle  \nonumber \\
K_{-}|n;k\rangle &=&\sqrt{n(2k+n-1)}|n-1;k\rangle  \nonumber \\
K_0|n;k\rangle &=&(n+k)|n;k\rangle
\end{eqnarray}
This is the discrete representation of SU(1,1) Lie algebra with Bargaman
index $k=(M+1)/2$. We see that the generalized Holstein-Primakoff
realization gives rise to the representation of SU(1,1) on the space $S$.
Note that $|M\rangle =|0;k\rangle $, the NBS can be written as 
\begin{equation}
|\eta ,M\rangle =e^{\xi (K_{+}-K_{-})}|0;k\rangle .
\end{equation}
This shows the NBS are essentially Peremolov's coherent states. 

\section{The NBS as excited geometric state, intermediate number-geometric
state and nonlinear coherent state}

\subsection{As excited geometric states and intermediate number-geometric
states}

From Eq.(10) we obtain 
\begin{eqnarray}
|\eta ,M\rangle &=&\eta ^{(M+1)/2}\sum_{n=0}^\infty {%
{M+n \choose M}%
}^{1/2}(1-\eta )^{n/2}|M+n\rangle  \nonumber \\
&=&\frac{\eta ^{(M+1)/2}}{\sqrt{M!}}a^{\dagger M}\sum_{n=0}^\infty (1-\eta
)^{n/2}|n\rangle  \nonumber \\
&=&\frac{\eta ^{M/2}}{\sqrt{M!}}a^{\dagger M}|\eta \rangle _g,
\end{eqnarray}
where 
\begin{equation}
|\eta \rangle _g=\eta ^{1/2}\sum_{n=0}^\infty (1-\eta )^{n/2}|n\rangle
\end{equation}
is the geometric state[16-21], which is also called Susskind-Glogower phase
state[11], phase eigenstate[16,17],and coherent phase state[19]. The photon
number distribution is $\eta (1-\eta )^n$, the geometric distribution. From
Eq.(18) the NBSs can be generated by repeated application of the creation
operator $a^{\dagger }$ on the geometric states. This shows that the NBS
belongs to an interesting class of nonclassical states, excited quantum
states. These states are first introduced by Agarwal and Tara as excited
coherent states[22]. So the NBSs can be viewed as excited geometric states.

Setting $M=0$ in Eq.(18), we get 
\begin{equation}
|\eta,0\rangle=|\eta\rangle_g,
\end{equation}
which shows that the NBS reduces to the geometric states for $M=0$. Note
that the NBS reduces to the number state $|M\rangle$ in the limit of $%
\eta\rightarrow 1$. Thus, the NBS interpolates between the number state and
geometric state and can be viewed as number-geometric state. 

\subsection{As nonlinear coherent states}

The geometric states are the eigenstates of the Susskind-Glogower phase
operator[23] $(1+N)^{1/2}a$, obeying the equation 
\begin{equation}
(N+1)^{-1/2}a|\eta \rangle _g=\sqrt{1-\eta }|\eta \rangle _g.
\end{equation}
Comparing with the definition of the nonlinear coherent states$|\alpha
\rangle _{nl}$[24,25] 
\begin{equation}
f(N)a|\alpha \rangle _{nl}=\alpha |\alpha \rangle _{nl},
\end{equation}
we know that the geometric states are nonlinear coherent states with the
nonlinear function $f(N)=(N+1)^{-1/2}$.

In a previous work, we have proved a general result that the excited
nonlinear coherent states are still nonlinear coherent states[26]. Since the
geometric states are nonlinear coherent states and the NBSs can be
recognized as excited geometric states, we infer that the NBSs are nonlinear
coherent states. In fact, multiplying Eq.(7) by the annihilation operator $a$
from the left , we get 
\begin{equation}
(N+1-M)a|\eta ,M\rangle =\sqrt{1-\eta }\sqrt{N+1-M}(1+N)|\eta ,M\rangle .
\end{equation}
Since we discuss the problem in the space $S$(Eq.(15)), we can multiply
Eq.(23) by $1/[(1+N)\sqrt{N+1-M}]$ from left. This leads to 
\begin{equation}
\lbrack \sqrt{N+1-M}/(N+1)]a|\eta ,M\rangle =\sqrt{1-\eta }|\eta ,M\rangle .
\end{equation}
The above equation shows that the NBS are nonlinear coherent states with the
nonlinear function $\sqrt{N+1-M}/(N+1)$. Eq.(24) naturally reduces to
Eq.(21) for $M=0$.

\section{Nonclassical properties}

\subsection{Sub-Poissonian distribution}

The simplest way to investigate the statistical characteristics of the
radiation field is to differentiate the generation function 
\begin{equation}
G(\lambda )=\sum_{n=0}^\infty P(n)\lambda ^n=\lambda ^M\left( \frac \eta {%
1+\lambda \eta -\lambda }\right) ^{M+1}
\end{equation}
with respect to the auxiliary real number $\lambda $. Here $%
P(n)=|C_n^{}(\eta ,M)|^2$ is the photon distribution function of the NBS.
The factorial moments are defined as $F(n)=\frac{d^nG}{d\lambda ^n}%
|_{\lambda =1}$. From Eq.(25) we obtain the factorial moments $F(1)$ and $%
F(2)$ as 
\begin{eqnarray}
F(1) &=&\langle N\rangle =\frac{M+1}\eta -1, \\
F(2) &=&\langle N^2\rangle -\langle N\rangle =\frac{(M+2)(M+1)}{\eta ^2}-4%
\frac{M+1}\eta +2.
\end{eqnarray}
Then we can easily derive Mandel's Q parameter 
\begin{eqnarray}
Q &=&\frac{\langle N^2\rangle -\langle N\rangle ^2-\langle N\rangle }{%
\langle N\rangle }  \nonumber \\
&=&\frac{F(2)-F^2(1)}{F(1)}=\frac{\eta ^2-2(M+1)\eta +M+1}{\eta (M+1-\eta )},
\end{eqnarray}
which measures the deviation from the Poisson distribution which corresponds
to the coherent state with $Q=0$. If $Q<0(>0)$, the field is called
sub(super)-Poissonian. The denominator of Eq.(28) is positive since $\eta
\le 1$, while the numerator can be positive or negative. For $M=0$, $%
Q=(1-\eta )/\eta \ge 0$. The NBS $|\eta ,0\rangle $(geometric state) is
super-Poissonian except $\eta =1$. For $M>0$, the condition for the
numerator $\eta ^2-2\eta (M+1)+M+1<0$ is 
\begin{equation}
\eta >\eta _{-}=M+1-\sqrt{M(M+1)}.
\end{equation}
It can be proved that $0<\eta _{-}<1$. Thus the NBS $|\eta ,M\rangle
^{-}(M>0)$ is super-Poissonian when $\eta <\eta _{-}$ and sub-Poissonian
when $\eta >\eta _{-}$. As $M$ increases, $\eta _{-}$ decreases and the
sub-Poissonian range increases. 

\subsection{Squeezing effect}

Define the quadrature operators $X$(coordinate) and $Y$ (momentum) by 
\begin{equation}
X=\frac 12(a+a^{\dagger }),Y=\frac 1{2i}(a-a^{\dagger }).
\end{equation}
Then their variances 
\begin{equation}
Var(X)=\langle X^2\rangle -\langle X\rangle ^2,\langle Var(Y)=\langle
Y^2\rangle -\langle Y\rangle ^2
\end{equation}
obey the Heisenberg's uncertainty relation 
\begin{equation}
Var(X)Var(Y)\ge \frac 1{16}.
\end{equation}
If one of the $Var(X)$ and $Var(Y)$ is less than 1/4, the squeezing occurs.
In the present case, $\langle a\rangle $ and $\langle a^2\rangle $ are real.
Thus, the variances of $X$ and $Y$ can be written as 
\begin{eqnarray}
Var(X) &=&\frac 14+\frac 12(\langle a^{\dagger }a\rangle +\langle a^2\rangle
-2\langle a\rangle ^2), \\
Var(Y) &=&\frac 14+\frac 12(\langle a^{\dagger }a\rangle -\langle a^2\rangle
).
\end{eqnarray}
From Eq.(2), the expectation values $\langle a\rangle $ and $\langle
a^2\rangle $ are obtained as 
\begin{eqnarray}
\langle a\rangle &=&\eta ^{M+1}(1-\eta )^{-M}\sum_{n=M}^\infty {%
{n+1 \choose M}%
}^{1/2}{%
{n \choose M}%
}^{1/2}(1-\eta )^{n+1/2}(n+1)^{1/2} \\
\langle a^2\rangle &=&\eta ^{M+1}(1-\eta )^{-M}\sum_{n=M}^\infty {%
{n+2 \choose M}%
}^{1/2}{%
{n \choose M}%
}^{1/2}(1-\eta )^{n+1}[(n+2)(n+1)]^{1/2}
\end{eqnarray}
Using Eqs.(26), and (33)-(36), we can investigate the squeezing effect.

By numerical calculations, we find that the squeezing occurs in both the
quadrature $X$ and $Y$. Fig.1 gives the variance of the quadrature $X$
versus $\eta $ for different $M$. It can be seen that the range and degree
of squeezing increase as $M$ increases. There exists a critical value of $M$%
. When $M<7$, there is no squeezing for arbitrary values of $\eta $. The
squeezing also occurs in the quadrature $Y$ as shown in Fig.2. In contrary
to the squeezing in the quadrature $X$, the range and degree decrease as $M$
increases. When $M$ is larger than a critical value 31, no squeezing occurs.

\section{The Q and Wigner functions}

If a field is prepared in a quantum state described by a density operator $%
\rho $, we can define the $s$-parametrized quasiprobability distribution in
phase space as[27] 
\begin{equation}
P(\beta ,s)=\frac 1{\pi ^2}\int d^2\xi C(\xi ;s)\exp (\beta \xi ^{*}-\beta
^{*}\xi ),
\end{equation}
where the quantum characteristic function is 
\begin{equation}
C(\xi ,s)=Tr[D(\xi )\rho ]\exp (s|\xi |^2/2).
\end{equation}
Here $\beta =x+iy$, with $(x,y)$ being the $c$ numbers corresponding to the
quadratures $(X,Y)$, and $D(\xi )=\exp (\xi a^{\dagger }-\xi ^{*}a)$ is
Glauber's displacement operator. It is possible to write the $s$%
-parametrized quasiprobability distribution as a infinite series [28] 
\begin{equation}
P(\beta ,s)=\frac 2\pi \sum_{n=0}^\infty (-1)^k\frac{(1+s)^k}{(1-s)^{k+1}}%
\langle \beta ,k|\rho |\beta ,k\rangle ,
\end{equation}
where $|\beta ,k\rangle =D(\beta )|k\rangle $ is the so-called displaced
number state. The expression above is suitable for direct numerical
calculations.

\subsection{Q function}
If we take $s=-1$, Eq.(39) is reduced to the familiar expression for the $Q$
function 
\begin{equation}
Q(\beta )=\frac 1\pi \langle \beta |\rho |\beta \rangle .
\end{equation}
Note that since the NBS can be viewed as an excited geometric state [see
Eq.(18)], we obtain the Q function of the NBS as 
\begin{equation}
Q(\beta )=\eta ^{M+1}\exp (-|\beta |^2)|\beta |^{2M}\left| \sum_{n=0}^\infty
\beta ^n(1-\eta )^{n/2}/\sqrt{n!}\right| ^2/M!.
\end{equation}
In Fig.3 we present plots of the $Q$ function of a NBS for different values
of $\eta $ and $M=5$. We can clearly see the deformation of the $Q$
function. When $\eta =1$, the $Q$ function representing the number state $%
|M=5\rangle $ is formed[Fig3.(d)] as expected. From the $Q$ function we can
also study the squeezing effects by examining the deformation of their
contours. Fig.4 is the contour plot of the $Q$ functions of two particular
NBSs, with (a) $\eta =0.2,M=5$ and (b) $\eta =0.85,M=50$. We can clearly see
the compression along the $y$ and $x$ direction, which corresponds to
squeezing in the $Y$ quadrature of the first NBS and in the $X$ quadrature
of the second. 

\subsection{Wigner function}

By taking $s=0$ in Eq.(39), we obtain a series representation for the Wigner
function 
\begin{equation}
W(\beta )=\frac 2\pi \sum_{k=0}^\infty (-1)^k\langle \beta ,k|\rho |\beta
,k\rangle .
\end{equation}
Now we insert Eq.(1) into the above equation, which yields 
\begin{equation}
W(\beta )=\frac 2\pi \sum_{k=0}^\infty \left| \sum_{n=M}^\infty C_n^{}(\eta
,M)\chi _{nk}(\beta )\right| ^2.
\end{equation}
In the expression above, the matrix elements $\chi _{nk}(\beta )=\langle
n|D(\beta )|k\rangle $ are given by[29] 
\begin{equation}
\chi _{nk}(\beta )=\beta ^n(-\beta ^{*})^k\exp (-|\beta
|^2/2)_2F_0(-n,-k;|\beta |^{-2})/\sqrt{n!k!},
\end{equation}
where $_2F_0(\alpha ,\beta ;z)$ are the generalized hypergeometric
functions[30]. The present form of $\chi _{nk}(\beta )$ is convenient for
numerical calculations.

The Wigner function can be used to trace the nonclassical behaviors of
quantum states. It is known that the negativity of the Wigner function is a
sufficient but not necessary condition for having nonclassical effects. In
Fig.5 we give plots of the Wigner function of a NBS by numerical
calculations of Eq.(43) for different values of $\eta$ and $M=1$. As in
Fig.5(a), the negative part of the Wigner function is already noticeable for 
$\eta=0.3$. For $\eta=0.5$[Fig.5(b)], the negative part is pronounced. In
fig5(c), for $\eta=0.9$, the negative part is even larger, and finally, in
Fig.7(d), we have the full Wigner function of a number state $|1\rangle$($%
\eta=1$). The Wigner function becomes more and more negative as $\eta$
increases.

\section{Generation of the new NBS}

Let us discuss the dynamical generation of the NBS. We consider two
different methods. The first is quite straightforward in concept from the
displacement operator formalism(Eq.(13)) but might not be very easy to
achieve experimentally. The Hamiltonian is given by 
\begin{eqnarray}
H &=&H_0+i\chi [\sqrt{\hat{N}-M}a^{\dagger }\exp (-i\omega t)-a\sqrt{\hat{N}%
-M}\exp (i\omega t)]  \nonumber \\
H_0 &=&\omega a^{\dagger }a.
\end{eqnarray}
The constant $\chi $ is the coupling strength. The coupling is of
intensity-dependent type and is similar to those in some 
intensity-dependent Jaynes-Cummings models [31-34]. The unitary time
evolution operator in the interaction picture is 
\begin{equation}
U(t)=\exp [\chi t(\sqrt{\hat{N}-M}a^{\dagger }-a\sqrt{\hat{N}-M})].
\end{equation}

Supposing the system is initially prepared in the number state $|M\rangle $,
we find the system at time $t$ is the NBS 
\begin{equation}
U(t)|M\rangle =|1-\tanh ^2(\chi t),M\rangle .
\end{equation}
The second method of the generation of the NBS is based on the fact that the
NBS is the excited geometric state. The geometric state can be prepared in
the non-degenerate three-wave interaction system[35]. We can also generate
the geometric state by the non-degenerate parametric amplifier described by
the two-mode Hamiltonian[36] 
\begin{eqnarray}
\tilde{H} &=&\tilde{H}_0+i\chi [a_1^{\dagger }a_2^{\dagger }\exp (-2i\omega
t)-a_1a_2\exp (2i\omega t)],  \nonumber \\
\tilde{H}_0 &=&\omega _1a_1^{\dagger }a_1+\omega _2a_2^{\dagger }a_2,
\end{eqnarray}
where $a_1$ and $\omega _1$ ($a_2$ and $\omega _2$) are the annihilation
operator and frequency for the signal(idler) mode. Frequencies $\omega _1$
and $\omega _2$ sum to the pump frequency, $2\omega =\omega _1+\omega _2$.
The coupling constant $\chi $ is proportional to the second-order
susceptibility of the medium and to the amplitude of the pump. The unitary
time evolution operator in the interaction picture is 
\begin{equation}
\tilde{U}(t)=\exp [\chi t(a_1^{\dagger }a_2^{\dagger }-a_1a_2)]
\end{equation}

Suppose that the system is initially prepared in the state $|0,0\rangle
=|0\rangle _1\otimes |0\rangle _2.$ Then at any time $t$ the system is in
the state

\begin{equation}
|\eta \rangle _{tm}=\tilde{U}(t)|0,0\rangle =\eta ^{1/2}\sum_{n=0}^\infty
(1-\eta )^{n/2}|n,n\rangle 
\end{equation}
which is the two-mode geometric state in comparison with Eq.(19). Here $\eta
=1-\tanh ^2(\chi t).$

Once the two-mode geometric state is prepared, one can generate two-mode
NBS by the following procedure in analogy to that proposed by Agawal and
Tara[22]. Consider the passage of a two-level excited atom through a cavity.
Let the initial state of the atom-field system be $|\eta \rangle
_{tm}\otimes |e\rangle ,$ where $|e\rangle $ is the atomic excited state.
The interaction Hamiltonian has the form[37]

\begin{equation}
\bar{H}=\hbar (gS^{+}a_1+g^{*}S^{-}a_1^{\dagger }),
\end{equation}
where $S^{\pm }$ are the psedospin operators of the atom and $g$ is the
coupling constant. Since $g$ is generally small, the state at time $t$ can
be approximated by

\begin{equation}
|\psi (t)\rangle \approx |\eta \rangle _{tm}\otimes |e\rangle
-ig^{*}ta_1^{\dagger }|\eta \rangle _{tm}\otimes |g\rangle ,
\end{equation}
which is valid for interaction times $gt<<1.$ From the above equation we
observe that, if the atom is detected to be in the ground state $|g\rangle ,$
then the state of the field is reduced to $a_1^{\dagger }|\eta \rangle _{tm}.
$ An extension of the above arguments to the multiphoton processes would
imply that the state $a_1^{\dagger M}|\eta \rangle _{tm}$ can be produced in
multiphoton processes. For the multiphoton processes, the Hamiltonian
(Eq.(51)) is replaced by a new Hamiltonian with $a_1\rightarrow a_1^M.$
Thus, the above procedure for a multiphoton process will result in the state

\begin{equation}
|\eta ,M\rangle _{tm}=\eta ^{(M+1)/2}\sum_{n=0}^\infty {%
{M+n \choose M}%
}^{1/2}(1-\eta )^{n/2}|M+n,n\rangle .
\end{equation}
The above normalized state is just the two-mode NBS. In a short summary
two
methods are proposed to generate the NBS.

\section{Conclusions}

We have investigated the NBS induced recently by Barnett and found the
ladder operator formalism and displacement operator formalism of the NBS.
The algebra involved is SU(1,1) Lie algebra via the generalized
Holstein-Primakoff realization. We found that the NBS are essentially
Peremolov's coherent states.

As excited quantum states, the NBSs are excited geometric states. As
intermediate states, they interpolate between the number and geometric
states. According to the definition of the nonlinear coherent states, we
find that the NBSs are nonlinear coherent states.

The NBS can be sub-Poissonian or super-Poissonian. There exists a critical
point $\eta _{-}$. The NBS is sub-Poissonian when $\eta >\eta _{-}$ and
super-Poissonian when $\eta <\eta _{-}$. The squeezing occurs in both the
quadrature $X$ and $Y$ with two critical values of $M=7$ and 31,
respectively. There is no squeezing occurs in the quadrature $X$ for $M<7$
and in the quadrature $Y$ for $M>31$ for arbitrary values of $\eta $. The $Q$
and Wigner functions are studied numerically. They show that the NBSs have
prominent nonclassical properties. We have proposed two methods of generation
of the new NBS.

In addition, the remarkable properties of the new NBS seem to suggest that
it deserves further attention from both theoretical and application sides of
quantum optics.

\vspace{1cm}

{\bf Acknowledgment}: One of the authors(Wang) thanks for the discussions
with Prof. H.C.Fu and help of Prof. C.P.Sun. The work is partially supported
by the National Science Foundation of China with grant number:19875008.

\end{document}